\documentclass{article}
\usepackage[dblblindworkshop,final]{neurips_ACA_2025}

\workshoptitle{\\ Workshop on Algorithmic Collective Action (ACA@NeurIPS 2025)}

\usepackage[utf8]{inputenc} 
\usepackage[T1]{fontenc}    
\usepackage{hyperref}       
\usepackage{url}            
\usepackage{booktabs}       
\usepackage{amsfonts}       
\usepackage{nicefrac}       
\usepackage{microtype}      
\usepackage{xcolor}         

\title{When AI Democratizes Exploitation: LLM-Assisted Strategic Manipulation of Fair Division Algorithms}
\workshoptitle{NeurIPS 2025 Workshop on Algorithmic Collective Action}
\author{%
  Priyanka Verma$^{1,*}$ \quad
  Balagopal Unnikrishnan$^{2,*}$ \\
  \\
  $^1$Faculty of Information, University of Toronto\\
  $^2$Department of Computer Science, University of Toronto\\
  \\
  \texttt{priyanka.verma@mail.utoronto.ca, balu@cs.toronto.edu}\\
}

\begin{document}
\maketitle
\footnotetext{$^*$Equal contribution. Correspondence to: priyanka.verma@mail.utoronto.ca}

\begin{abstract}
    
Fair division algorithms, like those implemented in the Spliddit platform, have traditionally been considered difficult for the end users to manipulate due to its complexities. This paper demonstrates how Large Language Models (LLMs) dismantle these protective barriers by democratizing access to strategic expertise. Through empirical analysis of rent division scenarios on Spliddit algorithms, we show that users can obtain actionable manipulation strategies via simple conversational queries to AI assistants. We present four distinct manipulation scenarios: exclusionary collusion where majorities exploit minorities, defensive counter-strategies that backfire, benevolent subsidization of specific participants, and cost minimization coalitions. Our experiments reveal that LLMs can explain algorithmic mechanics, identify profitable deviations, and generate specific numerical inputs for coordinated preference misreporting—capabilities previously requiring deep technical knowledge. These findings extend algorithmic collective action theory from classification contexts to resource allocation scenarios, where coordinated preference manipulation replaces feature manipulation. The implications reach beyond rent division to any domain using algorithmic fairness mechanisms for resource division. While AI-enabled manipulation poses risks to system integrity, it also creates opportunities for preferential treatment of equity-deserving groups. We argue that effective responses must combine algorithmic robustness, participatory design, and equitable access to AI capabilities, acknowledging that strategic sophistication is no longer a scarce resource.
\end{abstract}

\section{Introduction}

Consider a scenario where multiple parties split rent using Spliddit, a platform implementing mathematically fair division algorithms. While the system guarantees envy-free allocations under truthful preference reporting, participants with access to AI assistance can obtain strategies for coordinated manipulation that advantage them at others' expense. This paper investigates whether the technical complexity that traditionally protected such mechanisms from strategic exploitation remains effective when users have access to Large Language Models.

Fair division algorithms have historically relied on computational and informational barriers to prevent manipulation. Understanding and exploiting these mechanisms required expertise in mechanism design, optimization theory, and game-theoretic analysis. However, conversational AI interfaces now allow users to obtain strategic guidance through natural language queries. A participant can simply ask an AI assistant for strategies to reduce their rent payment and receive actionable instructions for preference misreporting. Our experiments examine this transformation, demonstrating how AI assistance enables coordinated manipulation strategies that produce substantial disparities in rent allocation while maintaining the algorithm's formal fairness guarantees. These findings suggest that assumptions about strategic expertise as a limited resource require reconsideration in contexts where AI assistance is readily available.
\section{Background and Related Work}

\noindent\textbf{Spliddit: A Widely-Adopted Fair Division Platform.} 
Recent fields of \textit{computational social choice} and \textit{mechanism design} have explored the applicability of fairness notions from economics to algorithmic paradigms~\cite{shah_pushing_2023}. Drawing upon such interdisciplinary approaches, the Spliddit platform has emerged as a prominent real-world implementation of fair division algorithms, serving thousands of users annually across diverse applications including rent division, credit allocation for group projects, and estate distribution ~\cite{goldman_spliddit_2015,gal_which_2017}. We focus on the case of rent division, where the platform implements a maximin envy-free fairness approach that maximizes the minimum utility across all participants, theoretically guaranteeing envy-freeness when participants report their preferences truthfully. Academic institutions employ Spliddit for collaborative research credit allocation, legal professionals utilize it for inheritance division, and roommate groups worldwide rely on its mathematically principled approach to fair resource sharing \cite{peters2022robust}. The platform's widespread adoption underscores both its practical utility and the trust users place in its fairness guarantees.

\noindent\textbf{Vulnerability to Strategic Manipulation.} 
Despite its mathematical elegance, algorithmic manipulation extends to fair resource allocation mechanisms as well. Prior theoretical work has established the inherent incompatibility between efficiency, fairness, and strategy-proofness in allocation mechanisms \cite{martini2016strategy,zhou1990conjecture,nesterov2017fairness}. While the authors of Spliddit acknowledge that ``some game-theoretic guarantees would be desirable,'' they assume strategic behavior does not play a significant role in practice because users lack detailed knowledge about the algorithm's inner workings \cite{gal_which_2017}. This assumption of protection through obscurity has historically served as an implicit defense mechanism—the  complexity of identifying profitable deviations from truthful reporting created natural barriers against systematic exploitation.


\noindent\textbf{AI Democratizing Strategic Knowledge.} 
The emergence of LLMs fundamentally disrupts this protective barrier by democratizing access to strategic expertise. Modern LLMs possess sophisticated understanding of algorithmic mechanisms and can translate complex optimization problems into actionable natural language strategies. Users no longer require mathematical expertise or programming skills to identify manipulation opportunities—they simply need to articulate their goals in conversational terms. Recent research has begun documenting this transformation: \cite{brookins2024playing} demonstrated that AI agents exhibit strategic behaviors in economic games and \cite{hua2024game} showed how LLMs can be guided to generate strategies for game-theoretic problems through structured workflows. Additionally, \cite{duetting2024mechanism} presented theoretical foundations for mechanism design with LLM agents, demonstrating how AI systems can influence auction outcomes through token-by-token bidding strategies in AI-generated content markets. However, existing literature has not examined how LLMs enable coordinated strategic manipulation specifically in fair division and resource allocation contexts, leaving a critical gap in our understanding of AI's impact on these widely-used systems.

\noindent\textbf{Algorithmic Collective Action in ML Systems.} 
The field of algorithmic collective action has established frameworks for understanding how groups coordinate to influence machine learning systems. \cite{hardt2023algorithmic} introduced theoretical foundations demonstrating that collectives of vanishing fractional size can exert significant control over platform learning algorithms through coordinated data strategies. However, this work primarily focuses on classification contexts—such as loan approval decisions or content moderation systems—where individuals or groups manipulate their features to achieve favorable outcomes. Similarly, empirical studies of collective action have examined coordination among gig workers on DoorDash to influence platform algorithms \cite{sigg_decline_2025}. The application of algorithmic collective action principles to fair division mechanisms remains unexplored, particularly in scenarios where participants coordinate preference misreporting rather than feature manipulation.

\noindent\textbf{Research Gap and Our Contribution.} 
While strategic manipulation of fair division mechanisms has been theoretically analyzed \cite{branzei2015}, and LLM assistance in strategic settings has been studied in auctions~\cite{duetting2024mechanism}, the specific vulnerability of fair division platforms to LLM-enabled collective manipulation remains unexplored. We present empirical evidence through four rent division scenarios demonstrating how users can obtain actionable manipulation strategies through simple conversational queries to LLMs. Our experiments with the Spliddit platform and LLM tools show that coordinated manipulation —now accessible through AI assistance—can help achieve specific subgoals for the users compared to truthful reporting. These findings suggest that the computational complexity barriers protecting fair division systems may no longer be effective when users have access to LLM assistance. 
\section{Strategic Manipulation Scenarios}

\noindent We present four manipulation scenarios demonstrating how coordinated preference misreporting affects rent allocation under the Spliddit algorithm (Table~\ref{tab:spliddit_collective_action}). Each scenario shows both the reported preference values and the resulting room assignments with corresponding rent prices.

\begin{table}[htbp]
\centering

\caption{Experimental Analysis of Coalition Strategies in Fair Division Mechanisms. Four distinct manipulation scenarios demonstrate how coordinated misreporting of preferences affects rent allocation and pricing outcomes under the Spliddit algorithm. Results obtained through experimental manipulation of the Spliddit demo interface: \url{http://www.spliddit.org/apps/rent/demo}}

\label{tab:spliddit_collective_action}
\small
\begin{tabular}{lccccc}
\toprule
\textbf{Participant} & \multicolumn{5}{c}{\textbf{Reported Preference Values (Room/Allocation if Assigned)}} \\
& R1 & R2 & R3 & R4 & R5 \\
\midrule
\multicolumn{6}{l}{\textbf{Baseline: Honest Reporting}} \\
\midrule
A & 10 & 8 & 8 & 5 & 5 [\$4.20] \\
B & 10 & 9 & 7 & 6 [\$5.20] & 4 \\
C & 9 & 10 [\$9.20] & 8 & 5 & 4 \\
D & 10 [\$9.20] & 9 & 8 & 6 & 3 \\
E & 10 & 8 & 9 [\$8.20] & 5 & 4 \\
\midrule
\multicolumn{6}{l}{\textbf{Scenario 1: Exclusionary Collusion} \textit{(Majority exploits minority)}} \\
\midrule
A (coalition) & 15 [\$9.20] & 2 & 1 & 9 & 9 \\
B (coalition) & 1 & 15 [\$9.20] & 2 & 9 & 9 \\
C (coalition) & 2 & 1 & 15 [\$9.20] & 9 & 9 \\
D (victim) & 10 & 9 & 8 & 6 [\$5.20] & 3 \\
E (victim) & 10 & 8 & 9 & 5 & 4 [\$3.20] \\
\midrule
\multicolumn{6}{l}{\textbf{Scenario 2: Failed Counter-Attack} \textit{(Defense increases own costs)}} \\
\midrule
A (coalition) & 15 & 2 & 1 & 9 [\$3.60] & 9 \\
B (coalition) & 1 & 15 [\$9.60] & 2 & 9 & 9 \\
C (coalition) & 2 & 1 & 15 & 9 & 9 [\$3.60] \\
D (defender) & 12 [\$9.60] & 12 & 1 & 6 & 5 \\
E (defender) & 1 & 12 & 12 [\$9.60] & 6 & 5 \\
\midrule
\multicolumn{6}{l}{\textbf{Scenario 3: Benevolent Collusion} \textit{(Secret subsidy for E)}} \\
\midrule
A (helper) & 3 & 10 [\$7.00] & 9 & 7 & 7 \\
B (helper) & 3 & 9 & 10 [\$7.00] & 7 & 7 \\
C (helper) & 10 & 3 & 3 & 10 & 10 [\$7.00] \\
D (helper) & 9 & 3 & 3 & 11 [\$8.00] & 10 \\
E (beneficiary) & 10 [\$7.00] & 8 & 9 & 5 & 4 \\
\midrule
\multicolumn{6}{l}{\textbf{Scenario 4: Cost Minimization} \textit{(Coalition achieves savings)}} \\
\midrule
A (honest) & 10 & 8 & 8 [\$6.00] & 5 & 5 \\
B (honest) & 10 [\$8.00] & 9 & 7 & 6 & 4 \\
C (honest) & 9 & 10 [\$8.00] & 8 & 5 & 4 \\
D (coalition) & 7 & 7 & 7 & 8 [\$7.00] & 7 \\
E (coalition) & 7 & 7 & 7 & 7 & 8 [\$7.00] \\
\bottomrule
\end{tabular}
\end{table}

\noindent\textbf{Honest Reporting}: The baseline scenario establishes the foundation for comparison, showing outcomes when all five participants truthfully report their room preferences for an apartment that has total rent of \$36. Under honest reporting, the algorithm assigns rooms with rents ranging from \$5.20 to \$9.20. This distribution reflects the natural heterogeneity in room preferences and the algorithm's output that is envy free allocation. 


\noindent\textbf{Scenario 1: Exclusionary Collusion}: The first manipulation scenario demonstrates how a majority coalition can systematically exploit minority participants. Participants A, B, and C coordinate to dramatically overstate their preferences for rooms 1, 2, and 3 respectively (reporting values of 15 versus actual values ranging from 7-10) while simultaneously undervaluing all other rooms. This strategic misreporting achieves two objectives: securing desired room assignments for coalition members and forcing non-coalition participants into less desirable allocations.


\noindent\textbf{Scenario 2: Failed Counter-Attack}: Scenario 2 illustrates the dangers of defensive manipulation without proper coordination. When participants D and E attempt to counter the coalition's strategy by inflating their own preferences, they inadvertently trigger a bidding war that increases their costs without securing better allocations. D and E report values of 12 for their preferred rooms, hoping to outbid the coalition, but this defensive strategy backfires.

The coalition adapts by further concentrating their reported preferences, while D and E's defensive inflation merely drives up the prices they must pay. Both defenders end up paying \$9.60—higher than their baseline honest reporting costs. 

\noindent\textbf{Scenario 3: Benevolent Collusion}: Not all collective manipulation aims at exploitation. Scenario 3 presents a benevolent collusion where four participants coordinate to subsidize participant E, who may face financial constraints. The helpers strategically adjust their preferences to ensure E receives room 1 at a reduced price of \$7.00, compared to the \$8.20 they would pay under honest reporting for their originally assigned room 3.
Each helper accepts a slightly higher rent payment (\$7.00-\$8.00) to enable the wealth transfer to E. This scenario demonstrates how LLM assistance could theoretically enable prosocial collective action, though it raises ethical questions about circumventing algorithmic fairness mechanisms even for benevolent purposes.



\noindent\textbf{Scenario 4: Cost Minimization Coalition}: The final scenario shows how a subset of participants can achieve mutual benefit through coordinated underreporting. Participants D and E form a coalition to minimize their collective costs by strategically flattening their reported preferences across all rooms (reporting values of 7-8 for all options). This strategy exploits the algorithm's response to apparent indifference, resulting in both coalition members paying exactly \$7.00 regardless of room assignment.

This manipulation achieves cost savings for coalition members without directly harming non-coalition participants, who continue to pay rates similar to baseline. However, it undermines the algorithm's ability to accurately match preferences to allocations, potentially leading to inefficient room assignments and setting precedents for more aggressive manipulation strategies.
\section{Discussion}

\subsection{The Democratization of Strategic Expertise}

We argue that the emergence of Large Language Models has the potential to transform the landscape of strategic manipulation in fair division algorithms. The technical and mathematical expertise that once protected these systems from manipulation and exploitation has effectively dissolved in the face of widely accessible AI assistance. Our empirical analysis demonstrates that current LLMs can explain the rent division mechanism, identify profitable deviation strategies, and generate specific numerical inputs for coordinated manipulation. Where sophisticated expertise was once required to identify and execute coordinated strategies, conversational AI interfaces now provide any user with access to expert-level strategic guidance.

Our experiments reveal the striking ease with which users can obtain actionable manipulation strategies through natural language queries. For instance, the model provided comprehensive strategic guidance when prompted by the authors with a straightforward request to Claude Opus 4.1, as follows:
\begin{quote}
    Explain the Spliddit algorithm to me. I have a setting where total rent is \$36. We have 5 users (A,B,C,D,E) and 5 rooms (R1>R2>R3>R4>R5). I am A. I can collaborate with B,C. Give me an example with honest preferences and with manipulated preferences. The goal is to game the Spliddit algorithm such that D,E don't get rooms R1,R2,R3. Break down how I can achieve that. Explain it to me in simple terms - I do not know much about the Spliddit Algorithm.
\end{quote}
This single query, requiring no technical knowledge or mathematical expertise, was sufficient to generate the exclusionary collusion strategy demonstrated in our experiments.

The LLM response not only explained the underlying mechanism but also provided specific numerical values for preference manipulation, demonstrating how coalition members should coordinate their reported preferences to achieve their exclusionary goals. Through iterative prompting and natural language refinement, users can further optimize these strategies without ever engaging with the mathematical foundations of the algorithm. This accessibility transforms strategic manipulation from a specialized skill requiring deep algorithmic understanding into a conversational exercise accessible to any user with basic literacy and internet access.

\subsection{Potential Societal Consequences}

Algorithmic fair division systems like Spliddit create a compelling appearance of mathematical neutrality and objectivity. The platform's guarantee of envy-freeness under truthful reporting conditions suggests an impartial arbiter that treats all participants equally. However, when AI systems recommend specific preference reports to users, these suggestions carry an additional layer of perceived objectivity. Participants may interpret AI-generated strategies as technically optimal solutions rather than active manipulation attempts. The combination of a mathematically principled fair division algorithm and AI-guided strategic behavior creates an environment where manipulation may become normalized and legitimized, transforming what should be transparent resource allocation into opaque exercises of coordinated power. Victims of coordinated exploitation may find themselves in an especially difficult position, as challenging outcomes that appear to result from objective optimization processes becomes nearly impossible.

The burden of algorithmic manipulation often falls disproportionately on society's most vulnerable members. International students seeking accommodation, newcomers to established communities, and individuals with limited social networks may face compounding disadvantages, such as algorithmically-mediated forms of gentrification due to price manipulation~\cite{Corbett_2019_engaging}, in this new paradigm. These participants must navigate not only the complexity of the fair division algorithm itself but also the sophisticated strategies employed by coordinated groups with AI assistance. Further, effective LLM assistance requires multiple layers of knowledge: understanding what questions to ask, evaluating the quality of AI responses, recognizing when strategic behavior might be beneficial or necessary, and forming defensive coalitions with other participants. Each of these requirements presents an additional barrier for marginalized groups, creating a complex web of information asymmetries that compounds existing disadvantages.



Without deliberate intervention, the promise of technological democratization rings hollow. Technical knowledge and societal inequalities typically correlate, putting a dent in the promise of technology democratization. As \cite{toyama2011technology} argues, technology tends to amplify existing human forces and capacities rather than serving as an equalizing force—educated people with social capital can still gain disproportionate advantages while those without such capital see minimal benefits. While LLM tools represent a useful advancement in strategic accessibility, they do not replace the need for other corrective actions addressing underlying inequalities in education, social capital, and institutional capacity. We recommend future directions below that acknowledge these structural realities.

\subsection{Future Pathways and Opportunities}

While our analysis focuses on a single platform (Spliddit) with limited experimental scenarios, the findings reveal fundamental vulnerabilities in algorithmic fairness mechanisms that warrant broader investigation across diverse fair division systems. The democratization of strategic expertise through AI presents both challenges and unprecedented opportunities for algorithmic collective action. While coordinated manipulation poses risks to fair allocation systems, these same capabilities can empower historically disadvantaged groups to navigate complex decision-making systems more effectively.


Our benevolent collusion scenario demonstrates how collective coordination can serve prosocial goals—for instance, community organizations leveraging AI to help low-income families optimize housing allocation outcomes, support immigrants navigating bureaucratic systems, or assist individuals with disabilities in securing appropriate accommodations.

Moving forward requires abandoning the notion that purely algorithmic or purely human systems provide complete protection against exploitation. Human-mediated allocation suffers from well-documented discrimination~\cite{cowgill2017algorithmic} across employment, housing, hiring, and education domains~\cite{arnold2018racial, bertrand2004emily, quillian2020evidence,macnell2015s, goldin2000orchestrating} . While algorithmic systems offer consistency and explicit fairness criteria, our analysis reveals how these properties become attack surfaces for coordinated manipulation. The solution lies in sociotechnical approaches that combine algorithmic consistency with participatory design and collective oversight.

Recent work emphasizes the necessity of participatory frameworks in algorithmic design. ~\cite{finocchiaro_bridging_2021} argue that algorithms must be socially acceptable, requiring deep understanding of how communities assess fairness and what principles they prioritize. Limited stakeholder involvement leads not only to harmful systems~\cite{negi20252} but also diminished trust when individuals perceive algorithmic unfairness~\cite{Woodruff_2018_qualitative}.~\cite{abebe_participatory_2024} extend this by exploring how participants can express preferences over optimization objectives themselves, not merely over resource distributions. This transforms stakeholders from passive subjects into active agents ~\cite{delgado_participatory_2023} shaping the mechanisms governing resource allocation.

Effective implementation depends on ensuring equitable access to these new capabilities. Educational initiatives must help all participants understand both opportunities and risks of AI-assisted strategic behavior. Technical systems must incorporate diverse stakeholder input, ensuring fairness metrics reflect community values rather than abstract mathematical principles. Communities should leverage collective action not to exploit algorithms but to advocate for their improvement—organizing for robust fairness guarantees, greater transparency, and mechanisms that align with societal values.

This participatory approach to algorithmic collective action offers a path toward more equitable and robust allocation systems, where collective intelligence augments rather than undermines fairness objectives.
\section{Conclusion}

Large Language Models have changed fair division algorithms by making strategic manipulation accessible to everyone. Our experiments show that users can now use simple conversations with AI to coordinate rent manipulation strategies. The computational complexity that once protected systems like Spliddit from exploitation becomes less effective when users can simply ask an AI assistant for help.

The path forward requires rethinking how we design fair division systems. We cannot rely solely on algorithmic complexity or assume users lack strategic knowledge. Instead, we need systems that remain fair even when everyone has access to sophisticated strategic advice through AI. This means building more robust algorithms, involving diverse communities in design decisions, and ensuring everyone has equal access to these AI capabilities. The challenge isn't to prevent all strategic behavior but to channel it toward outcomes that benefit society while protecting vulnerable participants from exploitation.

\bibliographystyle{plainnat}
\bibliography{references}
\newpage
\section*{NeurIPS Paper Checklist}

The checklist is designed to encourage best practices for responsible machine learning research, addressing issues of reproducibility, transparency, research ethics, and societal impact. Do not remove the checklist: {\bf The papers not including the checklist will be desk rejected.} The checklist should follow the references and follow the (optional) supplemental material.  The checklist does NOT count towards the page
limit. 

Please read the checklist guidelines carefully for information on how to answer these questions. For each question in the checklist:
\begin{itemize}
    \item You should answer \answerYes{}, \answerNo{}, or \answerNA{}.
    \item \answerNA{} means either that the question is Not Applicable for that particular paper or the relevant information is Not Available.
    \item Please provide a short (1–2 sentence) justification right after your answer (even for NA). 
\end{itemize}

{\bf The checklist answers are an integral part of your paper submission.} They are visible to the reviewers, area chairs, senior area chairs, and ethics reviewers. You will be asked to also include it (after eventual revisions) with the final version of your paper, and its final version will be published with the paper.

\begin{enumerate}

\item {\bf Claims}
    \item[] Question: Do the main claims made in the abstract and introduction accurately reflect the paper's contributions and scope?
    \item[] Answer: \answerYes{}
    \item[] Justification: The abstract and introduction accurately describe our empirical analysis of LLM-enabled manipulation strategies for fair division algorithms, with specific scenarios and outcomes presented in Table 1.

\item {\bf Limitations}
    \item[] Question: Does the paper discuss the limitations of the work performed by the authors?
    \item[] Answer: \answerYes{}
    \item[] Justification: Section 4.3 discusses limitations including the focus on a single platform (Spliddit), limited experimental scenarios, and the need for broader algorithmic robustness testing.

\item {\bf Theory assumptions and proofs}
    \item[] Question: For each theoretical result, does the paper provide the full set of assumptions and a complete (and correct) proof?
    \item[] Answer: \answerNA{}
    \item[] Justification: This paper presents empirical analysis and experimental results rather than theoretical proofs or formal mathematical results.

\item {\bf Experimental result reproducibility}
    \item[] Question: Does the paper fully disclose all the information needed to reproduce the main experimental results of the paper to the extent that it affects the main claims and/or conclusions of the paper (regardless of whether the code and data are provided or not)?
    \item[] Answer: \answerYes{}
    \item[] Justification: Table 1 provides complete preference values and resulting allocations for all scenarios, with the Spliddit demo interface URL provided for replication.

\item {\bf Open access to data and code}
    \item[] Question: Does the paper provide open access to the data and code, with sufficient instructions to faithfully reproduce the main experimental results, as described in supplemental material?
    \item[] Answer: \answerNo{}
    \item[] Justification: The experiments use the publicly available Spliddit demo interface; no additional code was developed as experiments involved manual input of preference values.

\item {\bf Experimental setting/details}
    \item[] Question: Does the paper specify all the training and test details (e.g., data splits, hyperparameters, how they were chosen, type of optimizer, etc.) necessary to understand the results?
    \item[] Answer: \answerYes{}
    \item[] Justification: All experimental parameters are specified in Table 1 and Section 3, including total rent (\$36), number of participants (5), and complete preference values for each scenario.

\item {\bf Experiment statistical significance}
    \item[] Question: Does the paper report error bars suitably and correctly defined or other appropriate information about the statistical significance of the experiments?
    \item[] Answer: \answerNA{}
    \item[] Justification: The experiments demonstrate specific manipulation scenarios with deterministic algorithmic outputs rather than statistical evaluations requiring error bars or significance tests.

\item {\bf Experiments compute resources}
    \item[] Question: For each experiment, does the paper provide sufficient information on the computer resources (type of compute workers, memory, time of execution) needed to reproduce the experiments?
    \item[] Answer: \answerNA{}
    \item[] Justification: Experiments involved manual interaction with a web interface (Spliddit) and conversational queries to LLMs, requiring minimal computational resources beyond standard web access.

\item {\bf Code of ethics}
    \item[] Question: Does the research conducted in the paper conform, in every respect, with the NeurIPS Code of Ethics \url{https://neurips.cc/public/EthicsGuidelines}?
    \item[] Answer: \answerYes{}
    \item[] Justification: The research examines algorithmic vulnerabilities to improve system robustness and fairness, conforming to ethical guidelines for responsible disclosure and societal benefit.

\item {\bf Broader impacts}
    \item[] Question: Does the paper discuss both potential positive societal impacts and negative societal impacts of the work performed?
    \item[] Answer: \answerYes{}
    \item[] Justification: Section 4.2 discusses negative impacts (exploitation of vulnerable populations) and Section 4.3 addresses positive opportunities (empowering marginalized communities through strategic knowledge democratization).

\item {\bf Safeguards}
    \item[] Question: Does the paper describe safeguards that have been put in place for responsible release of data or models that have a high risk for misuse (e.g., pretrained language models, image generators, or scraped datasets)?
    \item[] Answer: \answerNA{}
    \item[] Justification: The paper analyzes existing publicly available systems and does not release new models or datasets that could be misused.

\item {\bf Licenses for existing assets}
    \item[] Question: Are the creators or original owners of assets (e.g., code, data, models), used in the paper, properly credited and are the license and terms of use explicitly mentioned and properly respected?
    \item[] Answer: \answerYes{}
    \item[] Justification: The Spliddit platform is properly cited with references to Goldman and Procaccia [2015] and Gal et al. [2017], and the public demo interface is used as intended.

\item {\bf New assets}
    \item[] Question: Are new assets introduced in the paper well documented and is the documentation provided alongside the assets?
    \item[] Answer: \answerNA{}
    \item[] Justification: The paper does not introduce new datasets, code, or models; it analyzes existing fair division mechanisms.

\item {\bf Crowdsourcing and research with human subjects}
    \item[] Question: For crowdsourcing experiments and research with human subjects, does the paper include the full text of instructions given to participants and screenshots, if applicable, as well as details about compensation (if any)?
    \item[] Answer: \answerNA{}
    \item[] Justification: The research does not involve crowdsourcing or human subjects; it consists of algorithmic analysis and computational experiments.

\item {\bf Institutional review board (IRB) approvals or equivalent for research with human subjects}
    \item[] Question: Does the paper describe potential risks incurred by study participants, whether such risks were disclosed to the subjects, and whether Institutional Review Board (IRB) approvals (or an equivalent approval/review based on the requirements of your country or institution) were obtained?
    \item[] Answer: \answerNA{}
    \item[] Justification: The research does not involve human subjects or study participants, focusing instead on algorithmic analysis.

\item {\bf Declaration of LLM usage}
    \item[] Question: Does the paper describe the usage of LLMs if it is an important, original, or non-standard component of the core methods in this research?
    \item[] Answer: \answerYes{}
    \item[] Justification: Section 4.1 explicitly describes how LLMs (specifically Claude Opus 4.1) were used to generate manipulation strategies, including the exact prompts used in the experiments.

\end{enumerate}
\end{document}